\documentclass[useAMS,usenatbib]{mn2e}

\usepackage{color}
\usepackage{colortbl}
\usepackage{multirow}
\usepackage{dcolumn}
\usepackage{amsmath}
\usepackage{amssymb}
\usepackage{graphicx}
\usepackage{epsfig}
\usepackage{changebar}
\usepackage{natbib}
\usepackage{multirow}
\usepackage{subfigure}
\usepackage{txfonts}
\usepackage{relsize}

\usepackage{aalongtable}
\usepackage{longtable,lscape}
\usepackage[figuresright]{rotating}

\newcolumntype{d}[1]{D{.}{\cdot}{#1}}

\newcolumntype{.}{D{.}{.}{-1}}

\newcommand{\nh}{NH$_3$}
\newcommand{\nprime}{N$^{\prime}_{\mathrm{col}}$}
\newcommand{\ncol}{N$_{\mathrm{col}}$}

\newcommand{\Tk}{$T_{\mathrm{k}}$}
\newcommand{\Tex}{$T_{\mathrm{ex}}$}
\newcommand{\Tr}{$T_{\mathrm{r}}$}

\title[Mapping Ammonia in Star-Forming Regions]{Estimating column density from ammonia (1,1) emission in star-forming regions}
\author[L. K. Morgan et al.]{L. K. Morgan\thanks{E-mail:
lkm@astro.livjm.ac.uk}, T. J. T. Moore, J.Allsopp and D.J. Eden\\
Astrophysics Research Institute, Liverpool John Moores University, Twelve Quays House, Egerton Wharf, Birkenhead CH41 1LD \\ }
\begin{document}

\date{Accepted ??. Received ??; in original form ??}

\pagerange{\pageref{firstpage}--\pageref{lastpage}} \pubyear{2010}

\maketitle

\label{firstpage}

\begin{abstract}
We present a new, approximate method of calculating the column density of ammonia in mapping observations of the 23 GHz inversion lines. The temperature regime typically found in star forming regions allows for the assumption of a slowly varying partition function for ammonia. It is therefore possible to determine the column density using only the (J=1,K=1) inversion transition rather than the typical combination of the (1,1) and (2,2) transitions, with additional uncertainties comparable to or less than typical observational error.
  The proposed method allows column density and mass estimates to be extended into areas of lower signal to noise ratio.  We show examples of column density maps around a number of cores in the W3 and Perseus star-forming regions made using this approximation, along with a comparison to the corresponding results obtained using the full two-transition approach. We suggest that this method is a useful tool in studying the distribution of mass around YSOs, particularly in the outskirts of the protostellar envelope where the (2,2) ammonia line is often undetectable on the short timescales necessary for large area mapping.
\end{abstract}


\section{Introduction}
\label{sec:introduction}
  The usefulness of ammonia as a tracer of conditions in the interstellar medium (ISM) has been recognised for many years (e.g. \citealt{Ziurys1981,Martin1978,Cheung1968}). Ammonia is one of the most abundant molecular species in interstellar clouds, with typical observed molecular abundance ratios ranging from 10$^{-9}$ - 10$^{-7}$ (e.g. \citealp{Morgan2012,Friesen2009,Rosolowsky2008}). The excitation of the (1,1) rotational inversion transition of ammonia at 1.3 cm at typical molecular cloud temperatures (10 - 100 K) is overwhelmingly due to collisional processes \citep{Ho1983} and thus, above a critical density of $\sim$4$\times\ 10{^3}$ cm$^{-3}$ \citep{Maret2009}, is sensitive to the temperature of the region under observation. These densities are typical for star-forming regions (e.g. \citealt{Larson2003}).
  
The hyperfine structure of the \nh ~(1,1) transition allows the optical depth associated with the line to be determined from a single observation, while other molecular species require separate observations of an (often rare) isotopologue. Observations of the (2,2) transition also allow measurement of the rotational and kinetic temperature of the gas through the line ratios. The proximity of the (1,1) and (2,2) transitions in frequency means that the two can often be measured in the same instrumental bandpass (i.e. in a single observation), eliminating or greatly decreasing uncertainties arising from calibration and pointing errors.

  The combination of optical depth and temperature measurements allow the column density of molecular gas to be determined through commonly used methods \citep{Ho1983} and the assumption of LTE. Through this application, the distribution of molecular gas column density in the dense ISM may be mapped. Such mapping simultaneously determines the velocity dispersion of the region under investigation, providing virtually all of the physical information that may be determined in a single set of observations. For these reasons ammonia has become one of the most useful observational tools in modern studies of young stellar objects (YSOs), (e.g., \citealt{Morgan2010,Friesen2009,Rosolowsky2008}). The ramifications for the study of star formation include the determination of virial ratios (c.f. \citealt{Kirk2007}), the relative abundances of nitrogen- vs. carbon-bearing molecular species (e.g. \citealp{Tafalla2004}) and the relationship between the distribution of submillimetre continuum emission from dust to the emission from the associated molecular gas \citep{Morgan2012,Friesen2009,Rosolowsky2008}).

  Recent developments in instrumentation, such as the K-band Focal Plane Array (KFPA) at the Green Bank Telescope (GBT) in West Virginia, mean that ammonia emission may now be mapped over larger areas than were previously possible within reasonable time constraints. However, the maps resulting from such observations are likely to suffer from poor signal-to-noise ratios (SNRs) in comparison to pointed observations, away from the position of peak emission. The reasons for this are that ammonia observations of star-forming regions are often based upon previous submillimetre continuum emission maps and pointed observations are usually made towards suspected high column-density lines of sight, while mapping observations are probing fainter regions. Also, mapping observations are often time-consuming in comparison to single pointings and so shorter integration times per-pixel are often required.

\begin{figure}
\begin{center} 
\includegraphics*[width=0.5\textwidth]{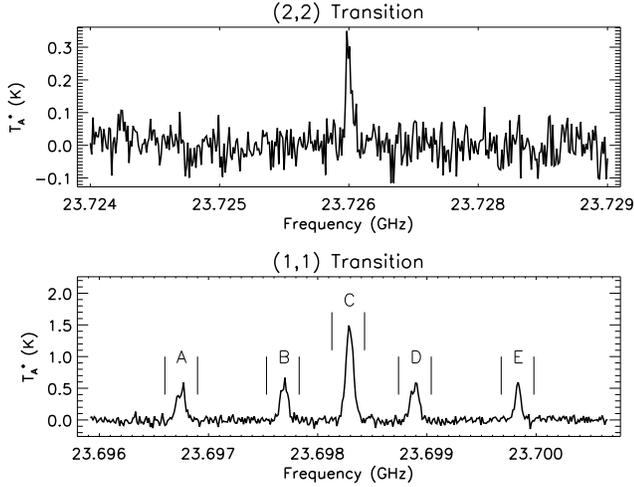}
\caption{Example \nh ~(1,1) (bottom) and (2,2) (top) spectra taken towards W3 at the coordinates 02:25:33.0 +61:13:21 (J2000)}
\label{fig:Spectrum}
\end{center}
\end{figure}

Recent results from large scale surveys such as the Red MSX Source (RMS) Survey and the Herschel infrared Galactic (HiGAL) Plane Survey indicate consistent average temperatures over a wide range of star-forming environments of $\sim$16 - 25 K \citep{Anderson2012,Urquhart2011,Elia2010,Stamatellos2010}. Over this temperature range the \nh ~(2,2) transition (corrected) antenna temperature ($T_{\mathrm{Am}(2,2)}^{*}$) is comparable to or greater than the line temperatures of the (1,1) hyperfine quadrupole transitions (often called `satellite' lines, labelled `A', `B', `D' and `E' in Fig. \ref{fig:Spectrum}).
  The outer (`A' and `E') and inner (`B' and `D') satellite quadrupoles have intensity ratios to the main quadrupole (`C') of 0.22 and 0.28 respectively \citep{Ho1983}. This corresponds to ratios of peak antenna temperature of $\sim$0.3 and, for a \mbox{main line optical depth ($\tau_{\mathrm{m} (1,1)}$)} of 1, $\frac{T_{\mathrm{Am}(2,2)}^{*}}{T_{\mathrm{Am}(1,1)}^{*}}$ ranges from 0.3 to 0.8. However, below the same temperature range of 16 - 25 K, $\frac{T_{\mathrm{Am}(2,2)}^{*}}{T_{\mathrm{Am}(1,1)}^{*}}$ falls below 0.3. Therefore, determination of optical depth, which requires detection of the (1,1) hyperfine quadrupole transitions, is likely to be ascertained over a greater spatial region than the rotational/kinetic temperature, which requires detection of the (2,2) main quadrupole. This limits the area over which column density may be determined to that in which the weaker (2,2) line is detected, despite more extensive detection of ammonia in the (1,1) transition.

In this paper we propose an approximate method of estimating column density along lines of sight in which only the (1,1) transition is detected.  The validity of the method requires only that the uncertainty introduced by variation of derived column density with kinetic temperature is less than typical observational errors over the range of temperatures found in star-forming regions.

\section{Method}
\label{sec:Method}
  We calculate the column density of ammonia molecules in the (1,1) transition following equation A4 of \citet{Friesen2009},
\begin{equation}\label{eq:coldensity}
 N(1,1) = \frac{8 \pi \nu _0 ^2} {\mathrm{c}^2 \mathrm{A}}  \frac{1+\mathrm{e}^{-\frac{\mathrm{h} \nu_{0}}{\mathrm{k_B} T_{\mathrm{ex}}}}}{1-\mathrm{e}^{-\frac{\mathrm{h} \nu_{0}}{\mathrm{k_B} T_{\mathrm{ex}}}}} \int \tau(\nu) d\nu,
\end{equation}

where $\nu _0$, \Tex\ and $\tau$ are the line rest frequency, measured excitation temperature and optical depth respectively. A is the Einstein spontaneous emission coefficient and h and $\mathrm{k_B}$ are the Planck and Boltzmann constants respectively. Here $\int \tau(\nu) d\nu = \sqrt{2 \pi} \sigma _V \left[\frac{\nu _0}{c}\right] \tau _{tot}$, where $\sigma _V$ is the measured line (velocity) dispersion and $\tau _{tot}$ is the total measured line optical depth, both determined through line fitting.

The total column density is determined by scaling the column density in the (1,1) state by the ratio of the overall partition function to the individual (1,1) partition function (Z/Z(J=1)), i.e. the probability of occupation of the (1,1) state.

  The partition function for the metastable (J = K) states of the ammonia molecule is
\begin{equation}
Z = (2\mathrm{J}+1) \mathrm{~S(J) ~exp}\frac{-\mathrm{h~[B~J~(J+1)+(C-B)~J^2}]}{\mathrm{k_B} T_\mathrm{k}}.
\end{equation}

  Here S(J) is taken as 2 for J=3, 6, 9 ... and 1 otherwise, accounting for the extra statistical weight of ortho-NH$_{3}$ ~over para-NH$_{3}$. The other terms in the partition function are, B and C - the rotational constants of the ammonia molecule, 298117.06 and 186726.36 MHz, respectively \citep{Pickett1998}; \Tk~is the kinetic temperature.

It can be seen then that the total column density of ammonia is dependent upon kinetic temperature (\Tk) in two ways, firstly through the contribution to the partition function and then in a more circuitous fashion as follows. In the assumption of LTE we have \Tex ~= \Tk, therefore, the population of the (1,1) state is also dependent upon \Tk. This is incorporated into our calculation of column density in equation \ref{eq:coldensity} in using \Tk ~as the temperature defining the relative population of states (from the Boltzmann distribution function). It should be noted that measurements of \Tex ~are often lower than \Tk, ~probably due to clumping of the gas on scales smaller than the telescope beam and the fact that \Tex ~is derived from the single (1,1) transition. This inequality means that it is usually not valid to use \Tex ~as an estimate of \Tk, particularly in single-dish observations.

\subsection{Temperature Range}
\label{subsec:Temp_Range}
Kinetic temperature is related to the rotation temperature through the following equation established by \citet{Walmsley1983} and updated with revised collisional rates by \citet{Swift2005}
\begin{equation}
T_\mathrm{r} = \frac{T_\mathrm{k}} {1+\frac{T_\mathrm{k}}{T_0}\ln\left[ 1+0.6\exp\left(-\frac{15.7}{T_\mathrm{k}}\right)\right]},
\end{equation}
Where $T_0 = \frac{E_{(2,2)} - E_{(1,1)}}{\mathrm{k_B}} \approx 41.5$K. It should be noted that, for \Tk\ $>$ $T_0$, \Tr\ may be significantly underestimated \citep{Walmsley1983}.\\

\Tr~is itself determined through the equation
\begin{equation}
 T_\mathrm{r} = \frac{-T_0}{\ln \left\{\frac{-0.282}{\tau_{\mathrm{m} (1,1)}} \ln \left[1-\frac{\Delta T_{\mathrm{Am}(2,2)}^{*}}{\Delta T_{\mathrm{Am}(1,1)}^{*}}\left(1-\mathrm{e}^{-\tau _{\mathrm{m}(1,1)}} \right) \right] \right\}},
\end{equation}
\citep{Ho1979}, where T$^{*}_{\mathrm{Am}}$ represents the measured (corrected) antenna temperature of the main line of the indicated transition and $\tau_{\mathrm{m}}$ is the optical depth of the main line.

It can be seen that the calculated rotational (and thus kinetic) temperature depends upon the ratio of antenna temperatures measured at the peaks of (1,1) and (2,2) line emission, in addition to the optical depth of the (1,1) transition. In order to determine the range of \Tk~for which we may wish to define column density in star-forming regions, we have examined the ammonia maps of Perseus and W3 presented in \citet{Morgan2012} in addition to other works.

  Measurement of the optical depth of the ammonia (1,1) transition requires detection of the hyperfine satellite quadrupoles (see Fig. \ref{fig:Spectrum}). The faintest of these is the quadrupole at $\sim$23.696 GHz, labelled `A' in Fig. \ref{fig:Spectrum} with an antenna temperature ratio relative to the main line of 0.26. As described above, deriving rotation (and hence kinetic) temperature from ammonia observations at 1.3 cm relies upon detection and measurement of the (2,2) main line. Therefore, if we are to determine the optical depth and rotation temperature from observations of the (1,1) and (2,2) ammonia lines then we require $\frac{T_{\mathrm{Am}(2,2)}^{*}}{T_{\mathrm{Am}(1,1)}^{*}} \geq 0.26$.

Measurements of the optical depth of ammonia in W3 and Perseus indicate that $\tau_{\mathrm{m}} <$ 5.0 in these regions \citep{Morgan2012}. However, an examination of previous observations indicate that it may be as high as 10.0 in some star-forming regions \citep{Rosolowsky2008,Longmore2007} (values of $\tau_{\mathrm{m}}$ may be higher than this in actuality but would likely be unobservable due to line saturation). The lowest values of $\tau_{\mathrm{m}}$ previously observed in star-forming regions are $\sim$0.1 \citep{Morgan2010,Wu2006} though it should be noted that values of $\tau_{\mathrm{m}}$ below $\sim$0.5 become difficult to measure through line strength ratios as the ratios approach the theoretical limit. Modelling and fitting of the hyperfine structure of the ammonia inversion transitions (e.g. \citealt{Morgan2012,Rosolowsky2008}) may provide lower values however.

\begin{figure}
\begin{center} 
\includegraphics*[width=0.45\textwidth]{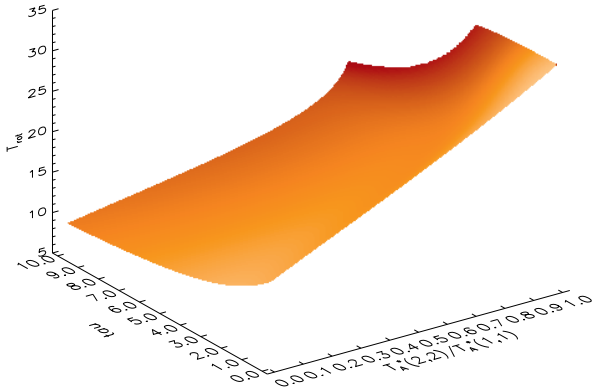}
\caption{Plane of the equilibrium solutions of \Tr~over typically observed ranges of $\frac{T_{\mathrm{Am}(2,2)}^{*}}{T_{\mathrm{Am}(1,1)}^{*}}$ and $\tau_{\mathrm{m}}$}
\label{fig:Tr_Plane}
\end{center}
\end{figure}

  Solving for \Tr~over the given ranges of $\frac{T_{\mathrm{Am}(2,2)}^{*}}{T_{\mathrm{Am}(1,1)}^{*}}$ and $\tau_{\mathrm{m}}$ allows us to identify a plane of solutions applicable to star-forming regions. The equilibrium solutions of this plane are shown in Fig. \ref{fig:Tr_Plane} and represent the range of values of \Tr~which may be expected to be found in any non-extreme star-forming region. \Tr~is found to vary between 8.7 and 32.8 K, corresponding to a kinetic temperature range of 8.9 - 45.5 K.

\section{Discussion}
\label{sec:Discussion}
  The various ways in which the assumed value of \Tk ~affects the calculated column density were addressed in section \ref{sec:Method}. The variation of column density over the expected kinetic temperature range described in Section \ref{sec:Method} (8.9 - 45.5 K) was examined for a grid of values of linewidth and optical depth (0.5 km/s $<$ dV $<$ 2.0 km/s, 0.1 $<$ $\tau$ $<$ 10.0) in order to investigate the effect of under- or over-estimating \Tk. The resulting change in column density was found to be as high as 80\% around the midpoint for any given set of dV and $\tau$ values.
  This uncertainty is relevant to the extreme case; within any given region observed values of \Tk\ typically vary by a smaller amount, e.g. for the Perseus low-mass star-forming region, values of \Tk\ range from $\sim$9K to $\sim$20K, while the W3 high-mass star-forming region exhibits a range of $\sim$12K $<$ \Tk $<$ $\sim$35K \citep{Morgan2012}. The standard deviations of the kinetic temperature distributions in these regions are 2.3 and 4.9 K for Perseus and W3 respectively. Using these standard deviations a more realistic 1$\sigma$ variation for the column density is $\sim$30\%
  
This uncertainty is equivalent to the measurement error in the estimation of optical depth, assuming an antenna temperature measurement error of 20\% which is appropriate for the observations used in this work and presented in \citet{Morgan2012}.

\begin{figure}
\begin{center} 
\includegraphics*[width=0.5\textwidth]{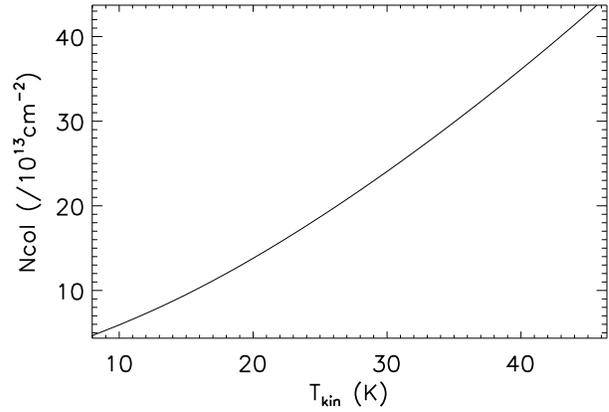}
\caption{The variation of ammonia column density with \Tk.}
\label{fig:Ncol}
\end{center}
\end{figure}

  The essentially small variation in the derived column density related to temperature means that column density can be calculated with the assumption of a mean temperature without any great loss of precision. This then allows column density to be determined for regions which do not have adequate SNRs in the (2,2) transition for kinetic temperature measurements.

\begin{figure}
\begin{center} 
\includegraphics*[width=0.5\textwidth]{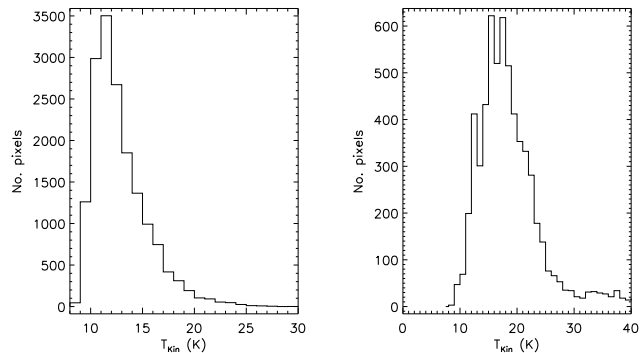}
\caption{The distributions of values of the kinetic temperature for maps of clumps in Perseus (left) and W3 (right).}
\label{fig:Tk_Hists}
\end{center}
\end{figure}

\citet{Jijina1999} presented a database compiled from the literature of 264 star-forming cores observed in the (1,1) and (2,2) lines of ammonia. They found a mean \Tk ~of 19 K, in agreement with values found in our own data and in other cited works. However, they note that this average temperature varies~from region to region and attribute this variation to association of sources with clusters, where clusters are associations of at least 30 embedded stars. Fig. \ref{fig:Tk_Hists} shows the distributions of values of \Tk ~derived from our KFPA maps in Perseus and W3, along lines of sight where both (1,1) and (2,2) detections were obtained. Ranges, averages and standard variations of the \Tk ~values for both regions are listed in Table \ref{tbl:rat_vals}. The plots show an approximately Gaussian distribution of \Tk ~values in W3 and an apparently skewed distribution in Perseus. It is noteworthy that the mean and median averages of the distribution in W3 are significantly higher ($\sim$17K) than for Perseus ($\sim$12K). It is possible that the distribution of values of \Tk ~is truncated in Perseus due to the lower average. Below temperatures of $\sim$9 K (2,2) to (1,1) line ratios fall below 0.3 and thus detection of the (2,2) line becomes increasingly difficult. Without a detection of the (2,2) line it is not possible to determine \Tk ~and so our sampling of \Tk ~values below 9 K is likely to be highly incomplete.

\subsection{Average-Temperature Column Density}
\begin{figure*}
\begin{center}
\includegraphics*[width=0.5\textwidth]{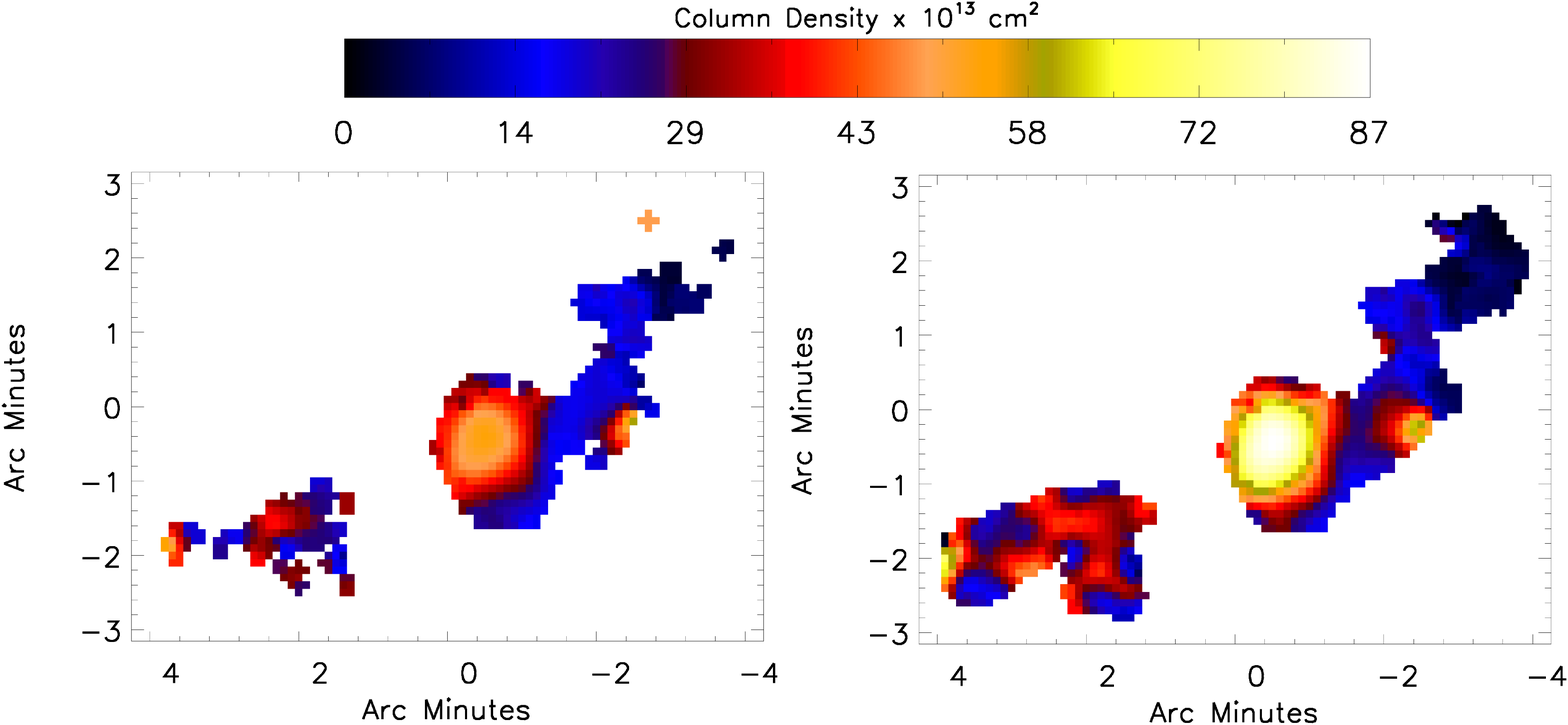}
\includegraphics*[width=0.5\textwidth]{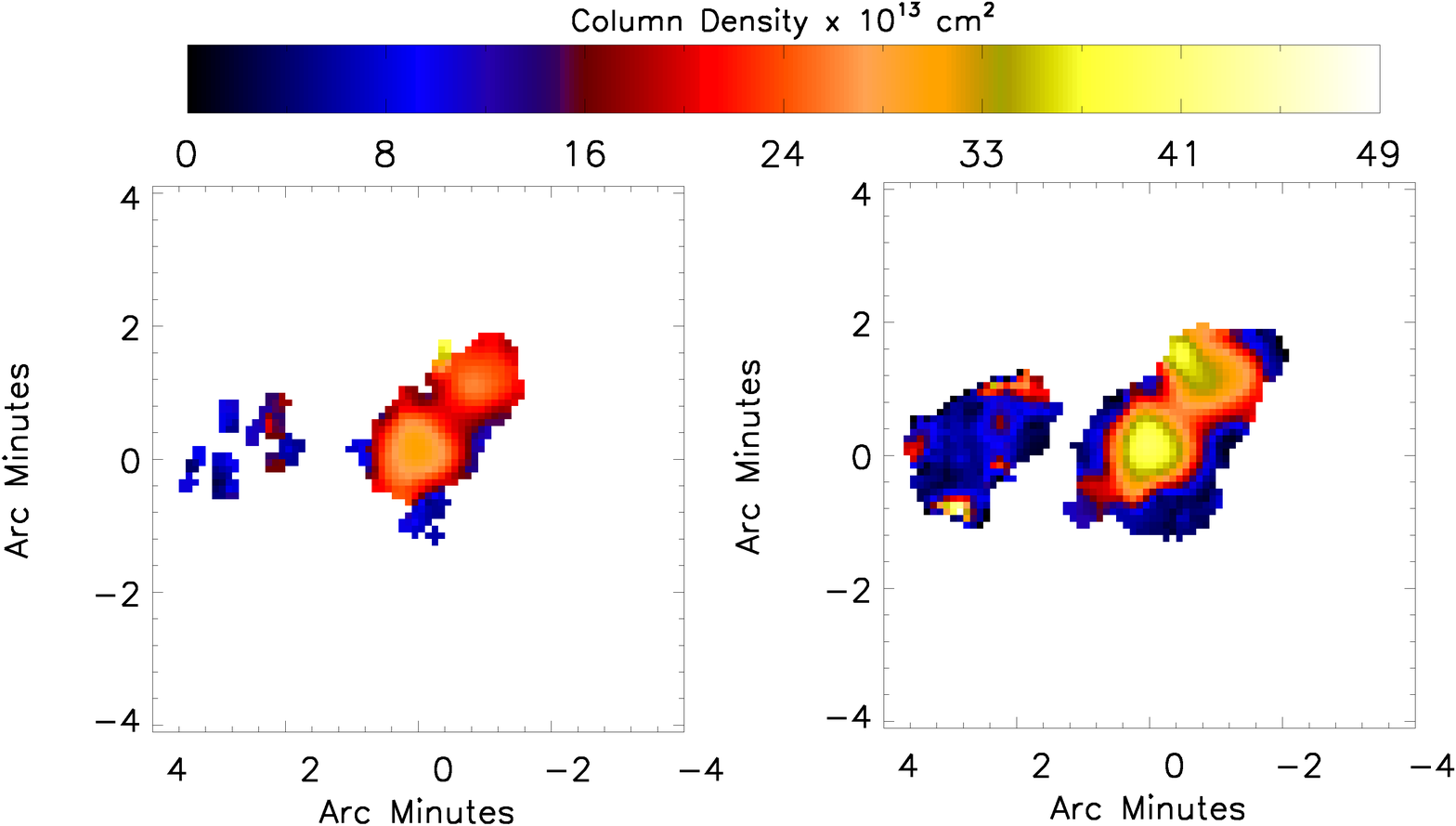}
\includegraphics*[width=0.5\textwidth]{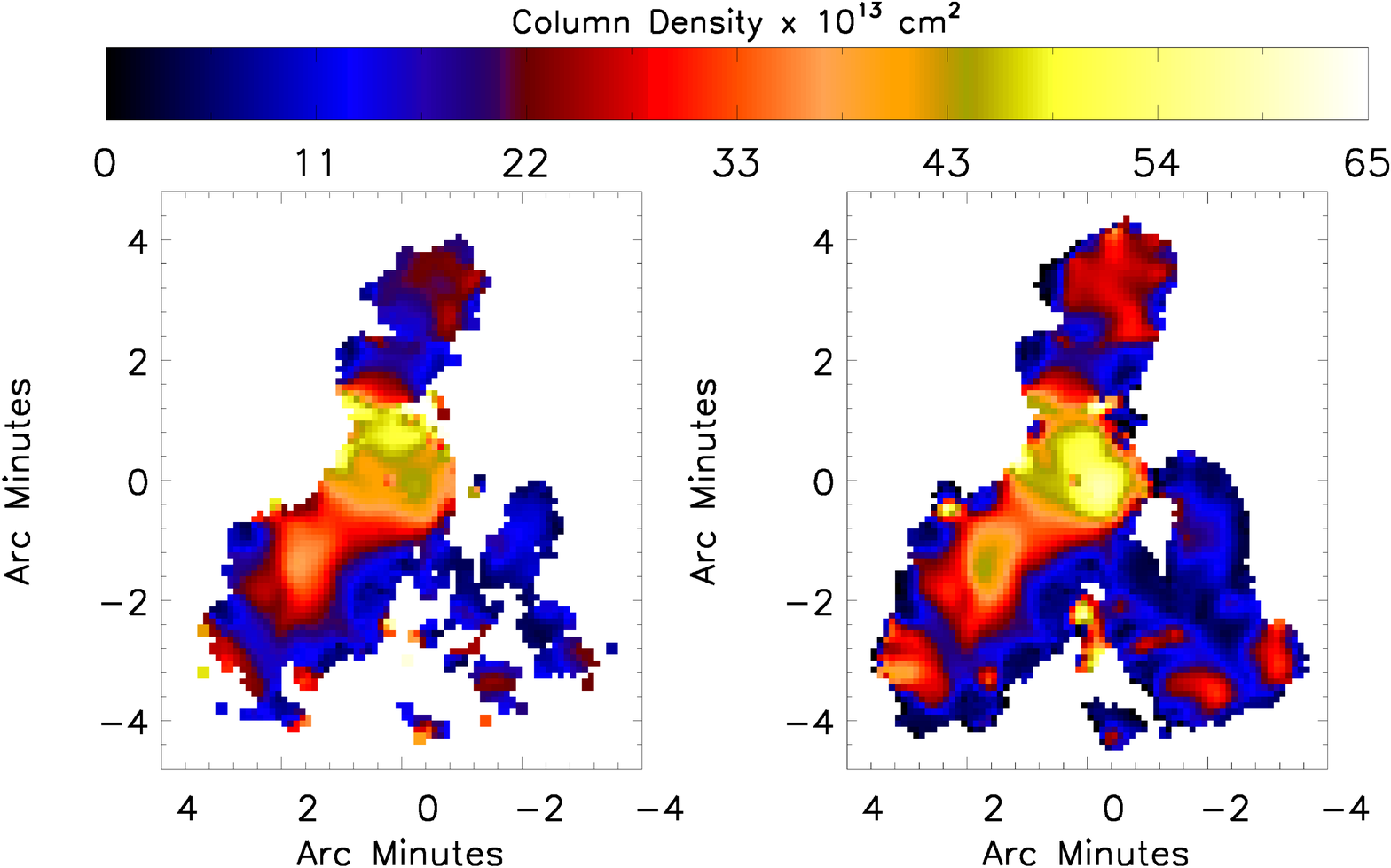}
\includegraphics*[width=0.5\textwidth]{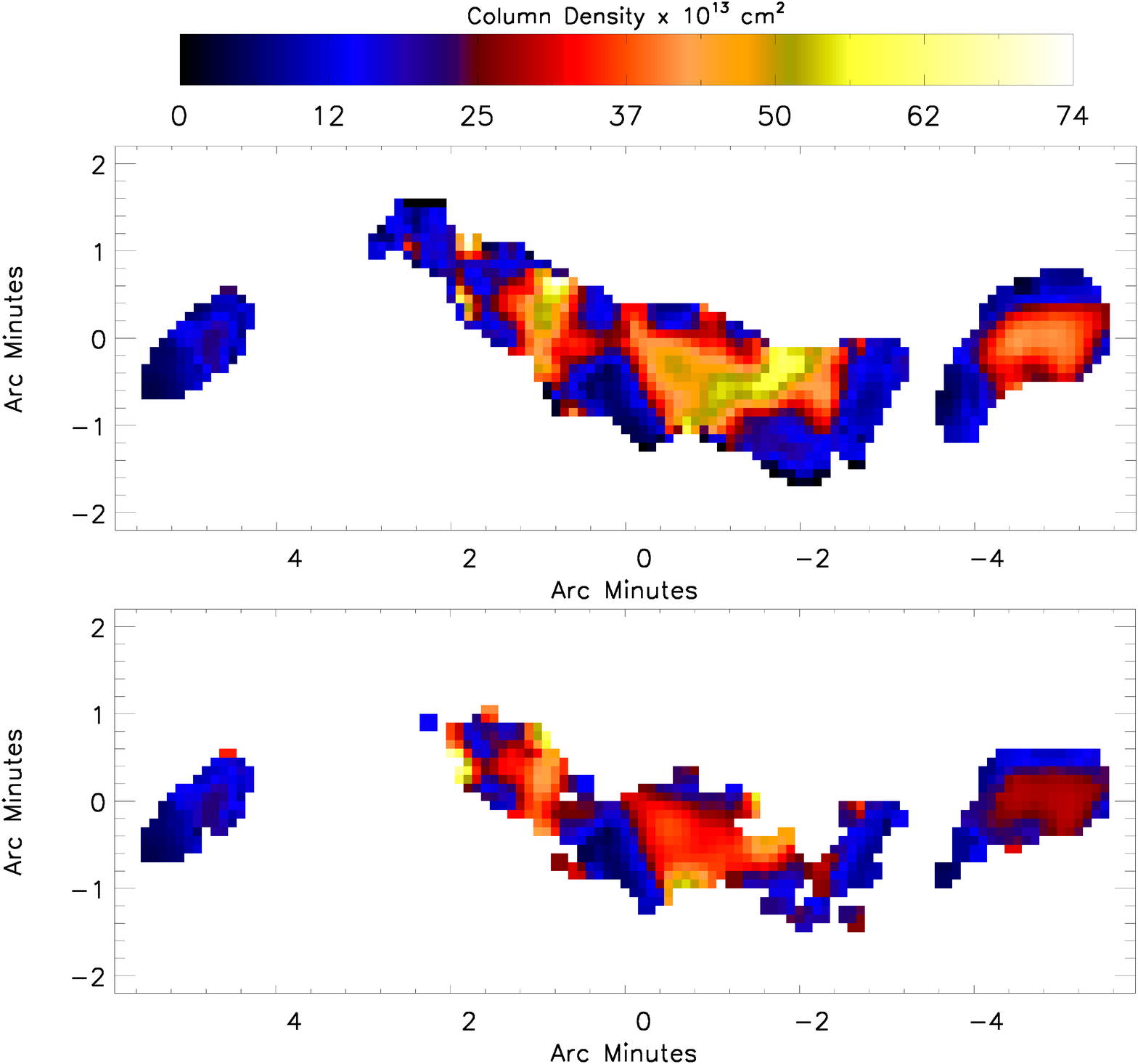}
\caption{Sections of the Perseus and W3 ammonia maps presented in \citet{Morgan2012} comparing the traditional determination of column density (left or bottom) to the described average-temperature determination (right or top).}
\label{fig:P147_comp}
\end{center}
\end{figure*}

We have estimated ammonia column densities from our KFPA data using both the traditional method and the proposed average-temperature method. The proposed average-temperature methods uses the mean kinetic temperature specific to the detected ammonia emission in each KFPA map. The results are presented in Fig \ref{fig:P147_comp}, where it can be seen that the area over which column density is determined increases by 20-30\% when using the average-temperature method.  This allows the column-density maps to extend to lower levels, with a minimum column density lower by a factor of $\sim$2 compared to the traditional determination of column density. The expansion of mapped column density, tracing the distribution of mass in dense regions, will be beneficial to star-forming studies. Accurate determinations of core sizes and tracing the radial mass distribution associated with ammonia cores is important in determining both virial ratios and the association of gas mass with dust mass. By minimizing the lowest known column density contour we can improve the quality and accuracy of such studies.

\begin{figure}
\begin{center} 
\includegraphics*[width=0.5\textwidth]{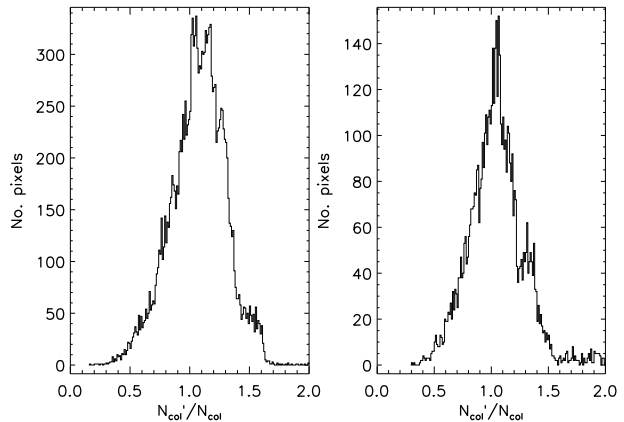}
\caption{The distributions of values of the average-temperature column density (N$^{\prime}_{\mathrm{col}}$) to traditionally-determined column density (N$_{\mathrm{col}}$) ratio. The values over mapped clumps in Perseus are shown on the left and values for mapped clumps in W3 are shown on the right.}
\label{fig:Rat_Hists}
\end{center}
\end{figure}

Where both the (1,1) and (2,2) lines were detected, we can compare the results of the two methods directly.  Fig \ref{fig:Rat_Hists} shows the distribution of ($\frac{\mathrm{N}^{\prime}_{\mathrm{col}}}{\mathrm{N}_{\mathrm{col}}}$), where N$^{\prime}_{\mathrm{col}}$ is the result of the approximate method and N$_{\mathrm{col}}$ of the exact method, along each line of sight, for both Perseus and W3.

It can be seen that \nprime ~does not diverge from \ncol ~by much more than 20\% over the majority of observed lines of sight and that the average-temperature method introduces an additional 1-sigma uncertainty of $\sim$ 25 - 30\%. Given that the uncertainties in \ncol ~due to measurement error are typically this size (see \citealt{Morgan2010}), the approximate method does not introduce significant additional error.

\begin{table}
\begin{center}
\caption{Statistics of the \Tk ~ and N$^{\prime}_{\mathrm{col}}$ to N$_{\mathrm{col}}$ Ratio distributions for Perseus and W3 clumps.}
\label{tbl:rat_vals}
\begin{minipage}{\linewidth}
\begin{center}
\begin{tabular}{lcccc}
\hline
\hline
{}			& \multicolumn{2}{c}{\Tk}	& \multicolumn{2}{c}{$\frac{\mathrm{N}^{\prime}_{\mathrm{col}}}{\mathrm{N}_{\mathrm{col}}}$}\\
{}			& Perseus	& W3	& Perseus	& W3 \\
\hline
Min			&  8.90		&  8.90	& 0.16  	&  0.30  \\
Max			& 31.00		& 51.50	& 9.28  	& 11.01  \\
Mean			& 12.84		& 18.73	& 1.07  	&  1.06  \\
Median			& 12.17		& 17.66	& 1.08  	&  1.04  \\
Std.Dev			&  2.65		& 5.86	& 0.25  	&  0.32  \\
\hline
\end{tabular}\\
\end{center}
\end{minipage}
\end{center}
\end{table}
As the average value of \Tk ~varies between different star-forming regions (Fig. \ref{fig:Tk_Hists}; \citealp{Jijina1999}) we can minimise the introduced error by adopting a mean \Tk ~value and range appropriate to the region under study. For example, an average \Tk ~determined from a catalogue of arbitrarily chosen star-forming cores may be representative of physical conditions on a Galactic scale. However, this may have little relevance to the column density of a particular region. In the presented examples the \Tk ~adopted was the mean over the individual mapped region (typically each KFPA map is circular with a radius of 5 arcminutes). In practice, the representative temperature which should be used will depend on the data at hand, although it has been shown (Section \ref{sec:Discussion}) that any introduced uncertainty is less than 50\% and probably less than 25\%.

   Given that the (2,2) transition of ammonia is easily simultaneously observable with modern instrumentation when making (1,1) observations, \Tk ~determinations should always be possible for at least part of an observed region. It is suggested then that use of the proposed method should not be used to replace knowledge of kinetic temperature but that knowledge of the distribution of \Tk ~for part of a given region may be used to expand the area over which column density is determined and probe a factor of two deeper.

\subsection{Systematic Effects}
It is possible that, by increasing the extent of column density mapping in star-forming regions using the described method, systematic biases on the values of measured column density may be introduced. There are two main causes of concern, firstly that the lower column density values being mapped with the temperature-averaging method actually represent lower volume density regions. Such regions may have significantly lower excitation temperatures due to the volume density being lower than the critical threshold, leading to systematic errors. Such an effect would likely be identified by a significant decrease in the optical depth of those regions. Values of optical depth below $\tau_{\mathrm{m} (1,1)} \sim$0.5 become difficult to measure using the line temperature ratio method described by \citet{Ho1983} and so values less than this can be used as a somewhat arbitrary limit, below which associated values of temperature-averaged column density might become suspect.
An examination of the optical depth values of the regions presented here indicate that this limit is not violated in any significant portion of the extended maps.

The other possible concern is that maps of column density using the proposed temperature-averaged method might be extended to the point that cores may be externally heated, by PDR heating in the external portions of the clouds for example. This effect has been observed by \citet{Morgan2009} in measuring the excitation temperatures of CO isotopologues at varying optical depths in star-forming regions close to HII regions. If this were to occur, then the assumption of an average temperature associated with a star-forming core would become invalid, again leading to systematic errors in the derivation of column density.

External heating is unlikely to be an issue in the determination of temperature-averaged column density for two reasons; firstly, it can be seen from Fig. \ref{fig:P147_comp} that the overall expansion of regions resulting from the assumption of an average temperature is small in comparision to the overall size of the regions. Thus, any effects of external heating would be evident in the observed values of \Tk\ and column-density determined in the usual manner. Secondly, the results of \citet{Morgan2009} show that C$^{18}$O temperatures are relatively independent of external heating and thus internal, super-critical density regions associated with low optical depth tracers are likely shielded or otherwise unaffected by significant external heating.

\section{Conclusion}
\label{sec:Conclusion}
The proposed method of deriving column density through the assumption of an average or representative temperature provides a way for the mass distribution in star formation regions to be traced over a significantly greater regime than previously possible. While this method cannot be expected to be a substitute for knowledge of the kinetic temperature, the ever-growing prevalence of, and need for, fast mapping of large regions means that it can be a useful tool. The accuracy of column density measurements made through the assumption of a mean temperature has been shown to be comparable to measurements made with full knowledge of temperature. Also, the additional uncertainty introduced is likely to be less than that from measurement errors.
These extended regions of emission are important to studies of star formation in many ways, not least in studies of the relation of submillimetre continuum emission to gas mass and the distribution of virial ratios in different regions \citep{Morgan2012}.\\

\noindent \textit{Acknowledgments}\\
The authors would like to thank an anonymous referee for a careful examination of this work which has resulted in a considerable improvement. We would also like to thank Erik Rosolowsky, James Urquhart and Charles Figura for useful relevant discussions.
We would like to thank the helpful staff of the Green Bank Telescope in the collection of data used in this paper related to the project GBT10C\_024. LKM is supported by a STFC postdoctoral grant (ST/G001847/1) and DJE is supported by a STFC PhD studentship. This research would not have been possible without the SIMBAD astronomical database service operated at CDS, Strasbourg, France and the NASA Astrophysics Data System Bibliographic Services.

\bibliography{References}

\bibliographystyle{mn2e}

\label{lastpage}

\end{document}